\documentclass[twocolumn,superscriptaddress,showpacs,preprintnumbers,amsmath,amssymb,nofootinbib]{revtex4-1}

\usepackage{latexsym}
\usepackage{graphicx}
\usepackage{dcolumn}

\newcommand{\vecI}{\mathrm{I}}
\newcommand{\bra}[1]{\langle #1 |}

\newcommand{\ket}[1]{| #1 \rangle}

\newcommand{\braket}[2]{\langle #1 | #2 \rangle}
\newcommand{\im}{\dot{\iota}\,}
\newcommand{\ketv}[1]{\ket{\mathbf{v}_{#1}}}
\newcommand{\brav}[1]{\bra{\mathbf{v}_{#1}}}
\newcommand{\ketvt}[1]{\ket{\widetilde{\mathbf{v}}_{#1}}}
\newcommand{\bravt}[1]{\bra{\widetilde{\mathbf{v}}_{#1}}}
\newcommand{\braketv}[2]{\langle \mathbf{v}_{#1} | \mathbf{v}_{#2} \rangle}
\newcommand{\ketu}[2]{\ket{\mathbf{u}_{#1}^{#2}}}
\newcommand{\brau}[2]{\bra{\mathbf{u}_{#1}^{#2}}}
\begin{document}

\title{High-fidelity entanglement purification using chains of atoms
       and optical cavities}

\author{Denis Gon\c{t}a}
\email{denis.gonta@mpl.mpg.de}
\affiliation{Institut of Optics, Information und Photonics,
             Friedrich-Alexander-University Erlangen-Nuremberg,
             Staudtstrasse 7, 91058 Erlangen, Germany}

\author{Peter van Loock}
\email{loock002@uni-mainz.de}
\affiliation{Institut of Physics,
             Johannes Gutenberg University Mainz,
             Staudingerweg 7, 55128 Mainz, Germany}

\date{\today}

\begin{abstract}
In our previous paper [Phys.~Rev.~A~\textbf{84}~042303~(2011)], we
proposed an efficient scheme to purify dynamically a bipartite entangled
state using short chains of atoms coupled to high-finesse optical cavities.
In contrast to conventional entanglement purification protocols, we avoid
controlled-NOT gates and thus reduce complicated pulse sequences and superfluous
qubit operations. In this paper, we significantly improve the output fidelity
of remotely entangled atoms by introducing one additional entanglement protocol
in each of the repeater nodes and by optimizing the laser beams required to
control the entire scheme. Our improved distillation scheme yields an almost unit 
output fidelity that, together with the entanglement distribution and swapping, 
opens an attractive route towards an efficient and experimentally feasible 
quantum repeater for long-distance quantum communication.
\end{abstract}

\pacs{03.67.Hk, 42.50.Pq, 03.67.Mn}

\maketitle

\section{Introduction}

In classical data transmission, repeaters are used to amplify the
data signals (bits) when they become weaker during their propagation.
In contrast to classical information, the above mechanism
is impossible to realize when the transmitted data signals carry
bits of quantum information (qubits). In an optical-fiber system, 
for instance, a qubit is typically encoded by a 
single photon which cannot be amplified or cloned without destroying
quantum coherence associated with this qubit \cite{nat299, pla92}.
Therefore, the photon has to propagate along the entire length
of the fiber, which causes an exponentially decreasing
probability to receive this photon at the end of the channel.

To avoid this exponential decay of a photon wave-packet and preserve 
its quantum coherence, the quantum repeater was proposed \cite{prl81}.
This repeater can be divided in three building-blocks which have to 
be applied sequentially. First, a large set of entangled photon pairs 
distributed over sufficiently short fiber segments are generated. The 
two subsequent steps, (i) entanglement purification \cite{prl76, prl77} 
and (ii) entanglement swapping \cite{swap}, are employed to extend
the short-distance entangled photon pairs over the entire length
of the channel. Using the entanglement purification, high-fidelity 
entangled pairs are distilled from a larger set of low-fidelity entangled 
pairs by means of local operations performed in each of the repeater 
nodes and classical communication between these nodes. The entanglement 
swapping, finally, combines two entangled pairs of neighboring segments
into one entangled pair, gradually increasing the distance of shared
entanglement.

Because of the fragile nature of quantum correlations and inevitable 
photon loss in the transmission channel, in practice, it poses a serious 
challenge to outperform the direct transmission of photons along the fiber. 
Up to now, only particular building blocks of a quantum repeater 
have been experimentally demonstrated, i.e., bipartite entanglement 
purification \cite{prl90, nat443}, entanglement swapping \cite{prl96, pra71},
and entanglement distribution between two neighboring nodes \cite{nat454, sc316}.
Motivated both by an impressive experimental progress and theoretical advances, 
moreover, various revised and improved implementations of repeaters and their
building-blocks have been recently proposed \cite{pra77, pra81, pra84a, rmp83, lpr}.

Practical schemes for implementing a quantum repeater are not
straightforward. The two mentioned protocols, entanglement purification
and entanglement swapping, in general, require feasible and reliable quantum 
logic, such as single- and two-qubit gates. Because of the high complexity 
and demand of physical resources, entanglement purification is the most
delicate and cumbersome part of a quantum repeater. The conventional
purification protocols \cite{prl77, pra59}, moreover, involve multiple
controlled-NOT (CNOT) gates which pose a serious challenge for most physical 
realizations of qubits, involving complicated pulse sequences and
superfluous qubit operations \cite{nat443, prl104, pra71a, prl85, pra78, pra67}.

\begin{figure*}[!ht]
\begin{center}
\includegraphics[width=0.95\textwidth]{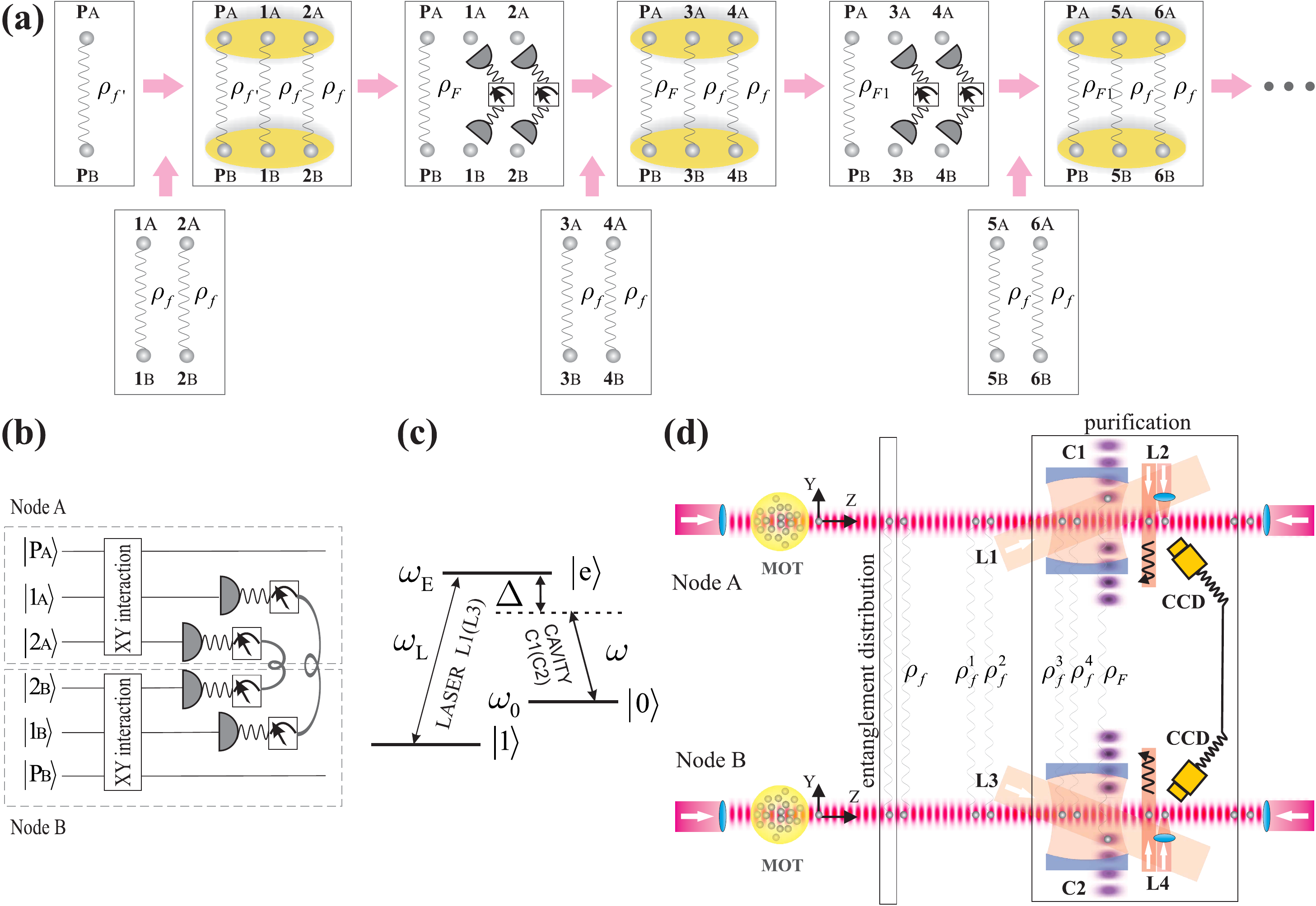} \\
\caption{(Color online) (a) Sequence of steps in the original
purification scheme. (b) Quantum circuit 
corresponding to the interaction indicated above by grey ellipses. 
(c) Structure of a three-level atom in the $\Lambda$-type configuration. 
(d) Experimental setup that realizes the purification scheme (a) and 
is incorporated into a quantum repeater segment with two neighboring 
nodes. See text for description.}
\label{fig1}
\end{center}
\end{figure*}

In our previous paper \cite{pra84}, we suggested a more practical scheme to purify
a bipartite entangled state by exploiting the natural evolution of spin chains
instead of CNOT gates. The realization of this dynamical scheme was proposed in 
the framework of cavity QED using short chains of atoms and optical cavities. 
In the present paper, we propose a modified purification scheme, in which we 
significantly improve the output fidelity of remotely entangled atoms. By 
introducing one additional entanglement protocol in each repeater node 
and by optimizing the laser beams required to control the entire scheme, we reach 
an almost unit output fidelity after the same number of purification 
rounds as before. This dramatic improvement, therefore, allows for multiple 
entanglement swapping operations and opens a route towards an efficient and 
experimentally feasible quantum repeater for long-distance quantum communication.

The paper is organized as follows. In the next section, we describe in detail 
the original purification scheme presented in our previous paper. In Sec.~III,
we present our modified high-fidelity purification scheme. We analyze the atomic 
evolution mediated by the cavity and laser field, and we determine the main properties 
which are relevant for our scheme in Sec.~III.A. In Sec.~III.B, we discuss 
a few relevant issues related to the implementation of our purification scheme,
while a short summary and outlook are given in Sec.~IV.

\section{Dynamical entanglement purification}

\begin{figure}[!ht]
\begin{center}
\includegraphics[width=0.4\textwidth]{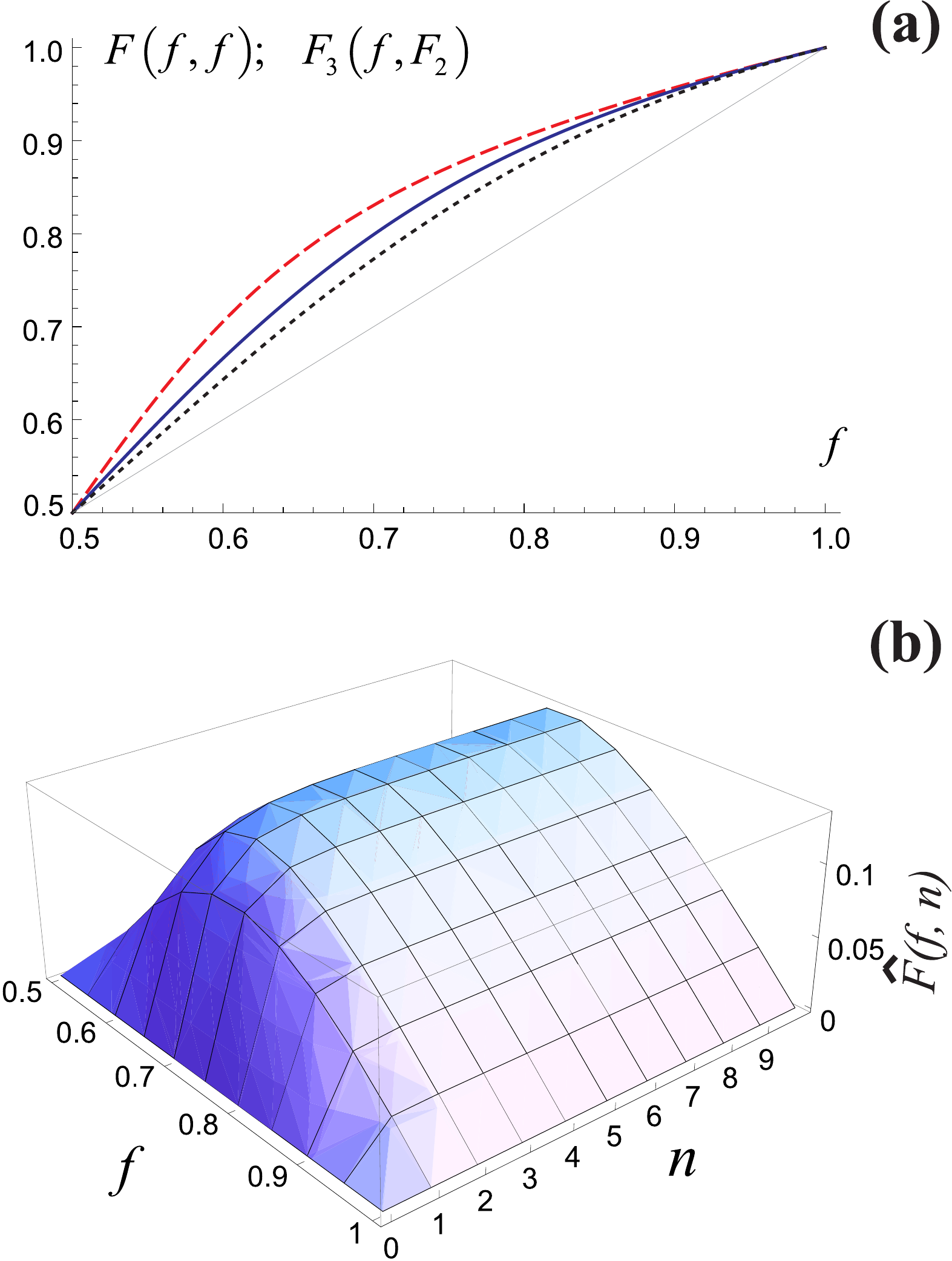} \\
\caption{(Color online) (a) Fidelities $F(f, f)$ (solid curve) and $F_3(f, F_2)$ 
(dashed curve) given by Eqs.~(\ref{final1}) and (\ref{final3}), respectively.
The dotted curve displays the fidelity given by Eq.~(34) in Ref.~\cite{pra84}.
(b) Plot of $\widehat{F}(f, n)$ given by Eq.~(\ref{seq4}) as a function of
input fidelity $f$ and number $n$ of purification rounds.}
\label{fig2}
\end{center}
\end{figure}

In our previously proposed purification scheme, two repeater nodes A and B
share one permanent qubit pair $P_{AB}$ and a finite set of temporary
(low-fidelity) entangled pairs grouped into elementary blocks of two qubit
pairs as displayed in Fig.~\ref{fig1}(a). Each temporary entangled pair is
given by the rank-two mixed state \footnote{In our previous paper,
we considered the Werner state $\rho_f = f \, \Phi^+ + \frac{1-f}{3}(\Phi^- +
\Psi^+ + \Psi^-)$ to describe low-fidelity entangled pairs. In this paper,
instead, we consider the state (\ref{density1}) that can be efficiently
generated between two remote nodes of a repeater using an optimal, ultimate 
entanglement distribution and detection protocol \cite{pra78b}. In the last 
section, we discuss this protocol and provide evidence supporting our choice.}
\begin{equation}\label{density1}
\rho_f = f \, \Phi^+_{A,B} + (1 - f) \Phi^-_{A,B},
\end{equation}
where $\Phi^\pm_{A,B} \equiv \ket{\phi^\pm_{A,B}} \bra{\phi^\pm_{A,B}}$ are
the Bell states in the \textit{qubit-storage} basis $\{ \ket{0}, \ket{1} \}$,
and where the fidelity
\begin{equation}\label{f-def}
\mathbf{F}(\rho_f) \equiv \text{Tr} \left[ \Phi^+_{A,B} \, \rho_f \right]
= f > 0.5
\end{equation}
is above the threshold value of $1/2$. The qubit-storage states
$\ket{0}$ and $\ket{1}$ correspond to the two long-living states of a three-level
atom in the $\Lambda$-configuration as displayed in Fig.~\ref{fig1}(c). In order
to protect this qubit against the decoherence caused by the fast-decaying excited
state $\ket{e}$, the states $\ket{0}$ and $\ket{1}$ are chosen as the stable ground
and long-living metastable states or as the two hyperfine levels of the ground state.

The permanent pair $P_{AB}$, characterized by the density operator $\rho_{f^\prime}$,
is supplemented by two temporary pairs $1_{AB}$ and $2_{AB}$, characterized by the
density operators $\rho_{f}^1$ and $\rho_{f}^2$, respectively, as seen Fig.~\ref{fig1}(a). 
Each of the repeater nodes $A$ and $B$, therefore, contains one triplet of qubits 
$P_A, 1_A, 2_A$ and $P_B, 1_B, 2_B$, respectively. Each of these triplets evolves due 
to the isotropic Heisenberg XY Hamiltonian \cite{ap16}
\begin{equation}\label{ham-xy}
H_{xy} = \frac{\hbar \, J_1}{2} \sum_{i = 1}^3
         \left( \sigma_i^x \sigma_{i+1}^x + \sigma_i^y \sigma_{i+1}^y \right),
\end{equation}
over the time period ($n = 0,1,2, \ldots$)
\begin{equation}\label{time}
T = \frac{\pi}{3} \left( n + \frac{1}{2} \right) J_1^{-1} \, ,
\end{equation}
where $\sigma^x_i$ and $\sigma^y_i$ are the respective Pauli operators 
in the \textit{cavity-active} basis $\{ \ket{0}, \ket{e} \}$, such that
$\sigma_4^x = \sigma_1^x$ and $\sigma_4^y = \sigma_1^y$, and where $J_1$ is
the coupling between the qubits. The above Hamiltonian
with periodic boundaries is produced deterministically in our scheme by coupling
of three (three-level) atoms to the same mode of a high-finesse resonator [see
Fig.~\ref{fig1}(c)]. In our previous paper, we identified the Hamiltonian
(\ref{ham-xy}) with the Jaynes-Cummings Hamiltonian in the large detuning limit,
i.e., $\Delta \gg g$, where $g$ is the atom-cavity coupling strength, $\Delta$ 
is the atom-cavity detuning, and $J_1 \equiv g^2 / \Delta$ is the coupling between 
the atoms subject to the same cavity mode.

The evolution governed by the Hamiltonian (\ref{ham-xy}) over the time period
(\ref{time}) is referred to below as the purification gate and is indicated by
the ellipses in Fig.~\ref{fig1}(a) and by rectangles in Fig.~\ref{fig1}(b).
Once the purification gate is performed, the qubit pairs $1_{AB}$ and $2_{AB}$
are pairwise projected in the computational (qubit-storage) basis
$\{ \ket{0}, \ket{1} \}$ and the outcome of these projections is exchanged between
the two nodes by means of classical communication [see Fig.~\ref{fig1}(b) and
the third box of Fig.~\ref{fig1}(a)]. Entanglement purification is successful
if the outcome of projections reads
\begin{equation}\label{outcome}
\{ 0_{1A}, 1_{2A}, 0_{1B}, 1_{2B} \} \quad \text{or} \quad
\{ 1_{1A}, 0_{2A}, 1_{1B}, 0_{2B} \} \, .
\end{equation}
In this case, the (unprojected) permanent qubit pair $P_{AB}$ is described
by the density operator
\begin{equation}\label{density2}
\rho_F = F(f, f^\prime) \, \Phi^+_{A,B}
                 + \left( 1 - F(f, f^\prime) \right) \Phi^-_{A,B} \, ,
\end{equation}
where
\begin{equation}\label{final}
F(f, f^\prime) = \frac{f^\prime -16 (f^\prime - 2) f + 32 (3 f^\prime - 1) f^2}
                         {81 + 32 f^2 - 80 f^\prime + 16 (10 f^\prime - 7) f} \, ,
\end{equation}
such that $\mathbf{F}(\rho_F) > \mathbf{F}(\rho_{f^\prime})$.
The entanglement purification is unsuccessful, if the mentioned
outcome of projections $1_{AB}$ and $2_{AB}$ disagrees with (\ref{outcome}).
In this case, the permanent pair $P_{AB}$ should be reinitialized and
the entire sequence from Fig.~\ref{fig1}(a) restarted.

The density operator (\ref{density2}) ensures that the (permanent)
qubit pair $P_{AB}$ preserves its rank-two form
after each successful purification round. Unlike the conventional
purification protocol, therefore, the purified state (\ref{density2})
is completely characterized by the fidelity
$\mathbf{F} \left( \rho_F \right) = F(f, f^\prime)$. The expression
(\ref{final}), furthermore, describes quantitatively how the input
fidelity $f^\prime$ of the permanent qubit pair is modified due to one 
single (and successful) purification round. In Fig.~\ref{fig2}(a), we 
compare the fidelity
\begin{equation}\label{final1}
F(f, f) = \frac{f (11 - 16 f + 32 f^2)}{27 - 64 f + 64 f^2}
\end{equation}
(solid curve) with the respective fidelity given by Eq.~(34) in Ref.~\cite{pra84}
(dotted curve) that was obtained within the same scheme, however, by
considering the Werner state instead of the rank-two mixed state
(\ref{density1}) in this paper. As expected due to the vanishing
contribution of $\Psi^\pm$ in (\ref{density1}), the growth of fidelity
in the case of a rank-two mixed state is larger as for the Werner state.

Assuming that each purification round is successful, the sequence from
Fig.~\ref{fig1}(a) leads to the gradual growth of entanglement fidelity (of
stationary atoms) with regard to the respective fidelity obtained in the
previous round
\begin{equation}\label{seq3}
f^\prime < F_{1} (f, f^\prime) < F_{2} (f, F_{1}) < \ldots < F_{n} (f, F_{n-1}).
\end{equation}
In order to understand how much the output fidelity increases with each
purification round, we analyze quantitatively the following sequence
\begin{equation}\label{seq4}
f < F_{1}(f, f) < \ldots < F_{n}(f, F_{n-1}) \equiv f + \widehat{F}(f, n) \, .
\end{equation}
In Fig.~\ref{fig2}(b), we show the plot of function $\widehat{F}(f, n)$ that
describes the difference between the final fidelity $F_{n}(f, F_{n-1})$ obtained
after $n$ (successful) rounds and the initial fidelity $f$ ($n = 0$). It is clearly
seen that during the first three rounds, this function exhibits a notably fast
growth that \textit{saturates} and, with increasing $n$, yields a negligible
growth with regard to the \textit{fixed point} fidelity
\begin{widetext}
\begin{equation}\label{final3}
F_{3}(f, F_{2}) = \frac{f \left( 70859 - 377904 f + 950112 f^2 - 1368064 f^3 +
1278976 f^4 - 671744 f^5 + 294912 f^6 \right)}{177147 - 1051072 f + 2792896 f^2 -
 4204544 f^3 + 3904512 f^4 - 2162688 f^5 + 720896 f^6} \, .
\end{equation}
\end{widetext}

Regardless of the number of (successful) purification rounds, therefore, the final
fidelity is bounded by a fixed point value that determines the optimal number of
purification rounds required to reach the best performance in a resource- and time-efficient way.
Such behavior is the common feature of the (so-called) \textit{entanglement pumping}
purification scheme introduced by W.~D\"{u}r and co-authors in Ref.~\cite{pra59}.
We refer to this property as the saturation of entanglement purification. Corresponding
to $n=3$ successful purification rounds, for which the final fidelity reaches its
saturation level, we display in Fig.~\ref{fig2}(a) the (fixed point)
fidelity $F_{3}(f, F_{2})$ by a dashed curve. By comparing this curve to the solid
curve, we conclude that the growth of fidelity due to three successive rounds
is notably larger if compared to the case of a single purification round.

\begin{figure*}[!ht]
\begin{center}
\includegraphics[width=0.95\textwidth]{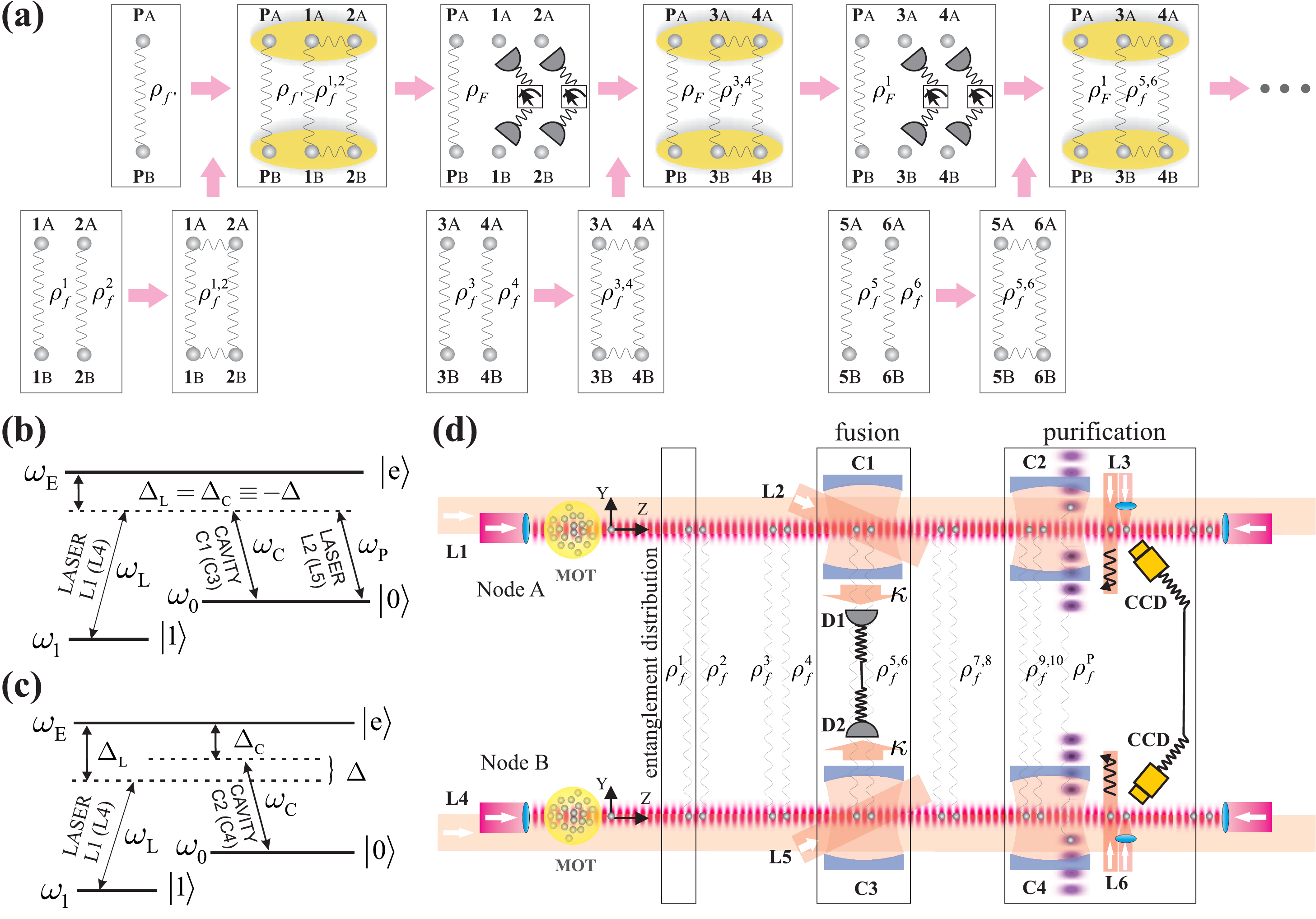} \\
\caption{(Color online) (Color online) (a) Sequence of steps in the modified
purification scheme. (b), (c) Structure of a three-level atom in the $\Lambda$-type 
configuration subjected to the cavity and laser fields. See text for description.
(d) Experimental setup that realizes the purification scheme (a) and 
is incorporated into a quantum repeater segment with two neighboring 
nodes.}
\label{fig3}
\end{center}
\end{figure*}

We remark that the described purification scheme is based on the effect
of entanglement transfer between the networks of evolving spin chains that
was introduced and investigated in Ref.~\cite{qip}. In the same reference,
it was suggested that this effect plays the key role in the entanglement
concentration once a part of the spins from two such networks are locally 
measured. One similar entanglement purification protocol, that is based on the 
natural spin dynamics, has been proposed independently in Ref.~\cite{pra78a}. 
In our scheme, the role of (spin-chain) networks is played by the atomic
triplets located in two repeater nodes, while the cavity-mediated interaction
governed by the Hamiltonian (\ref{ham-xy}) reproduces the spin-chain dynamics.
From a more fundamental point of view, the mentioned effect of entanglement
transfer originates the constructive and destructive interference of the
quantized spin waves (magnons) in an evolving spin chain (see \cite{pra78a} 
and references therein).

The main physical resources of the proposed purification scheme are:
(i) short chains of atoms, (ii) two high-finesse optical cavities, and (iii) detectors
for projective measurements of atomic states. In Fig.~\ref{fig1}(d) we show the experimental
setup of a quantum repeater segment that includes two neighboring nodes (A and B).
In this setup, each repeater node consists of one optical cavity $C_1$ ($C_2$) acting
along the $y$-axis, a laser beam $L_1$ ($L_3$), a chain of atoms transported by means
of an optical lattice along the same axis, one stationary atom trapped inside the cavity
with the help of a vertical lattice, laser beam $L_2$ ($L_4$) acting along the $y$-axis,
a magneto-optical trap (MOT), and a CCD camera connected to the neighboring node through
a classical communication channel.

We associated the permanent qubits with the stationary atoms trapped inside cavities $C_1$
and $C_2$, and the temporary qubits with (the chains of) atoms inserted into the horizontal
lattices and transported along the $z$-axis. According to the experimental scheme in
Fig.~\ref{fig1}(d), this identification implies that atoms pass sequentially through the
cavity, such that only two atoms from the chain couple simultaneously to the same cavity
mode. These two atoms together with the stationary (trapped) atom form an atomic triplet
in each repeater node as assumed by our purification scheme.

Right before an atom from node A enters the cavity, it becomes entangled with the
respective atom from node B as depicted in Fig.~\ref{fig1}(d) by wavy lines. This
entanglement is generated non-locally by means of an entanglement distribution block
(indicated by a rectangle), such that each produced entangled pair
is described by Eq.~(\ref{density1}) in the qubit-storage basis
$\{ \ket{0}, \ket{1} \}$. During the transition of an atomic pair through the cavity,
the triplet of atoms has to undergo the cavity-mediated evolution governed by the
Hamiltonian (\ref{ham-xy}) in each of the repeater nodes over the time period (\ref{time}).
Since the Hamiltonian acts solely on the cavity-active states $\{ \ket{0}, \ket{e} \}$,
the atomic population has to be mapped from the qubit-storage basis to the cavity-active
basis in order to make possible the interaction of atoms with the cavity mode and, moreover,
to protect the qubits against the decoherence caused by the fast-decaying excited state $\ket{e}$.
This mapping is realized using short resonant light pulses produced by the laser beam $L_1$
($L_3$). Each pulse transfers the electronic population from the qubit-storage states to
the cavity-active states (or backwards), such that the atoms couple to (or decouple from)
the cavity field in a controlled fashion.

According to the sequence in Fig.~\ref{fig1}(a), furthermore, the purification sequence
is completed once the states of an (conveyed) atomic pair are projectively measured and
the outcome of projections is pairwise exchanged between the repeater nodes in order to
decide if the purification was successful or not.
In our experimental scheme, the latter projections are performed by means of the laser beam
$L_2$ ($L_4$) and a CCD camera in each of the repeater nodes as displayed in Fig.~\ref{fig1}(d).
While the laser beam $L_2$ ($L_4$) removes atoms in a given (storage-basis) state from the
chain without affecting atoms in the other state (so-called push-out technique \cite{prl91}),
the CCD camera is used to detect the presence of remaining atoms via fluorescence imaging
and determine, therefore, the state of each atom that leaves the cavity.

In the successful case, furthermore, the next atomic pair is transported into the cavity
and the next purification round takes place with the same stationary atom (permanent qubit).
In the unsuccessful case, however, the stationary atoms have to be reinitialized and the
entire sequence from Fig.~\ref{fig1}(a) should be restarted.

The approach presented in this section requires that short atomic chains are transported 
with a constant velocity along the experimental setup and coupled to the cavity-laser fields
in a well controllable fashion. For this purpose, we introduced in our setup [see Fig.~\ref{fig3}(d)] 
(i) a magneto-optical trap (MOT) that plays the role of an atomic source and (ii) an optical 
lattice (conveyor belt) that transports atoms into the cavity from the MOT with a position 
and velocity control over the atomic motion. The proposed setup is compatible with existing
experimental setups \cite{prl95, prl98, njp10}, in which the above devices (i) and (ii) are
integrated into the same framework together with a high-finesse optical cavity. The 
number-locked insertion technique \cite{njp12}, moreover, enables one to extract atoms from 
the MOT and insert a predefined pattern of them into an optical lattice with a single-site 
precision. It was already demonstrated that an optical lattice preserves the coherence 
of transported atoms and can be utilized as a holder of a quantum register. By encoding the 
qubits by means of hyperfine atomic levels, a qubit storage time of the order of seconds
has been demonstrated within this register \cite{prl91, prl93}.

\section{High-Fidelity dynamical entanglement purification}

As seen from Fig.~\ref{fig2}(a), the output fidelity $F_3(f,F_2)$ (dashed curve)
obtained after three successful purification rounds is still far from unit
fidelity as required by a realistic quantum repeater. In fact, this output fidelity
enables one to perform only a few swapping operations between the purified entangled
pairs of neighboring repeater segments until the fidelity of the resulting pair (distributed
over a larger distance) drops to the initial fidelity $f$. Another bottleneck
in our scheme is the necessity to transfer the electronic population from the
qubit-storage states to the cavity-active states (and backwards) in order to
control the cavity-mediated evolution of atoms inside the cavity and protect our
qubits against the decoherence caused by the fast-decaying excited state $\ket{e}$
[see Fig.~\ref{fig1}(c)]. Obviously, these two obstacles make our scheme less attractive
to be considered in practice.

In this section, we propose a modified purification scheme, in which we significantly
improve the output fidelity of remotely entangled atoms and get rid of the superfluous
laser pulses required to transfer the electronic population of atoms. By introducing
one additional entanglement protocol in each repeater node and by optimizing
the laser beams required to control the entire scheme, we achieve an almost unit output
fidelity after the same number of successful purification rounds. This dramatic
improvement, therefore, allows for multiple entanglement swapping operations on the 
purified pairs.

Similar to the original scheme that we presented in the previous section, the modified
scheme includes two repeater nodes A and B sharing one permanent qubit pair $P_{AB}$,
characterized by the density operator $\rho_{f^\prime}$, and a finite set of temporary
entangled pairs as displayed in Fig.~\ref{fig3}(a). Each temporary entangled pair is
given by the rank-two mixed state (\ref{density1}) in the basis $\{ \ket{0}, \ket{1} \}$,
such that the fidelity (\ref{f-def}) of each pair is above the threshold value of $1/2$.
In contrast to the original scheme, however, right before the permanent pair is supplemented
by the temporary pairs $1_{AB}$ and $2_{AB}$, these two (separate) entangled pairs are
merged into the four-qubit entangled state
\begin{widetext}
\begin{eqnarray}\label{density3}
\rho_f^{1,2} &=& \frac{1}{2} \left( \ket{\phi^-_{1A,2A}, \phi^-_{1B,2B}}
                                 \bra{\phi^-_{1A,2A}, \phi^-_{1B,2B}} + \ket{\psi^-_{1A,2A}, \psi^-_{1B,2B}}
                                 \bra{\psi^-_{1A,2A}, \psi^-_{1B,2B}} \right) \notag \\
             &+& \frac{2f - 1}{2(1 - 2f + 2f^2)} \left( \ket{\phi^-_{1A,2A},
                     \phi^-_{1B,2B}} \bra{\psi^-_{1A,2A}, \psi^-_{1B,2B}} +
                 \ket{\psi^-_{1A,2A}, \psi^-_{1B,2B}} \bra{\phi^-_{1A,2A}, \phi^-_{1B,2B}} \right) \, .
\end{eqnarray}
\end{widetext}

This entangled state is generated using an additional entanglement protocol that
occurs prior to the purification gate in our scheme. By this protocol, the pairs
$1_A, 2_A$ and $1_B, 2_B$ interact locally within the repeater nodes $A$
and $B$, respectively, such that the state (\ref{density3}) is generated. In
Fig.~\ref{fig3}(d) we display the experimental setup of our modified scheme.
In contrast to the setup displayed in Fig.~\ref{fig1}(d), we added (i) high-finesse
cavities $C_1$ and $C_3$, (ii) photon detectors $D_1$ and $D_2$, and (iii) laser
beams $L_2$ and $L_5$ to each of the repeater nodes A and B, respectively. These
ingredients are compatible with the resources utilized in the original purification
scheme and they form together the \textit{fusion} block that is framed by a rectangle
in Fig.~\ref{fig3}(d). Finally, the laser beams $L_1$ and $L_4$ act continuously
along the z-axis and together with the cavity field of $C_1$ ($C_2$) and $C_3$
($C_4$), respectively, produce the two-photon (Raman) transition between the
states $\ket{0}$ and $\ket{1}$ of the coupled atoms, such that the fast-decaying
excited state $\ket{e}$ remains almost unpopulated (see below).

\begin{figure}[!t]
\begin{center}
\includegraphics[width=0.4\textwidth]{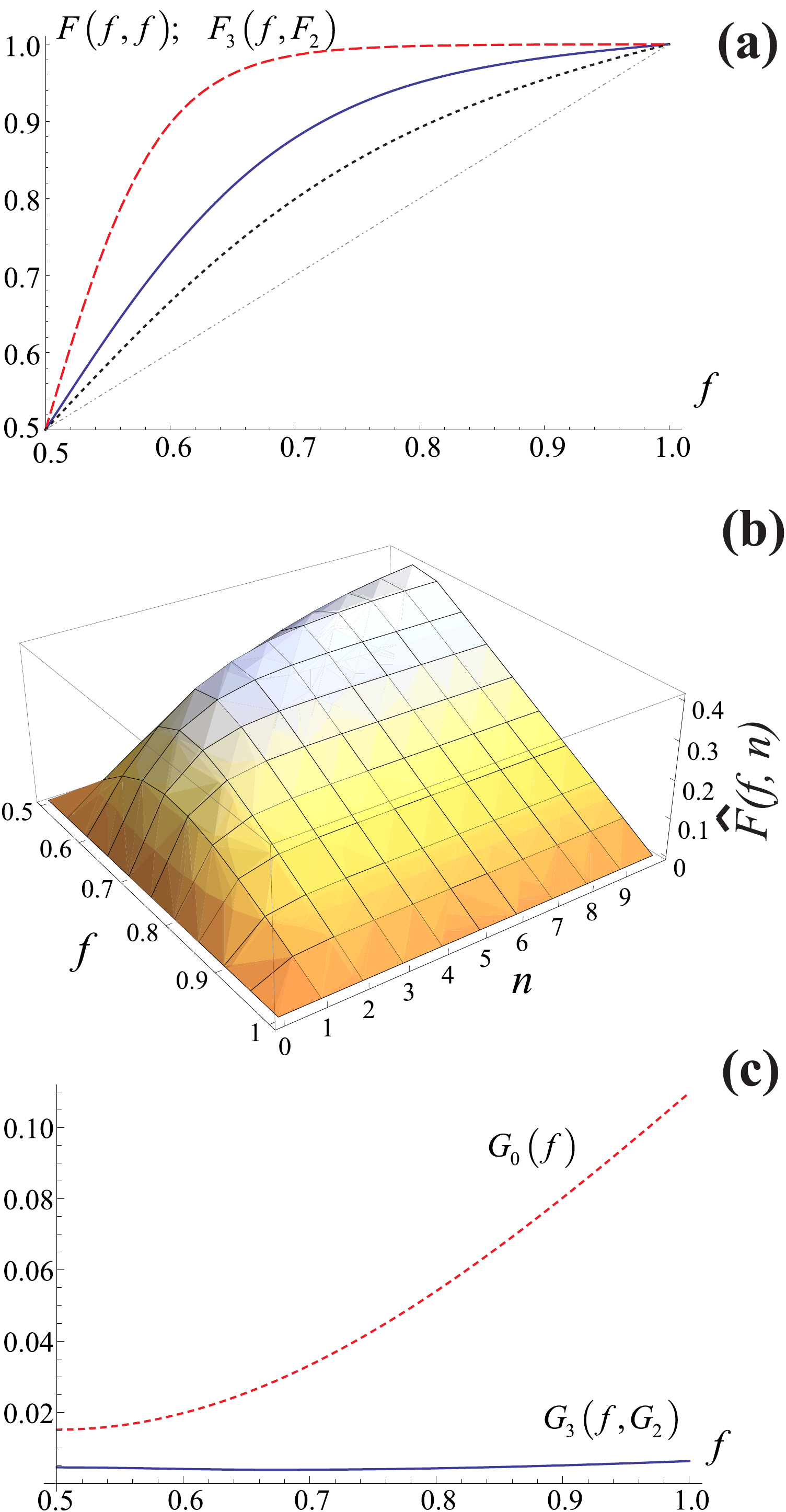} \\
\caption{(Color online) (a) Fidelities $F(f, f)$ (solid curve) 
and $F_3(f, F_2)$ (dashed curve) given by Eqs.~(\ref{final5}) and (\ref{final6}), 
respectively. The dotted curve displays the fidelity given by Eq.~(\ref{final1}) 
obtained in the original scheme. (b) Plot of $\widehat{F}(f, n)$ in the
modified scheme as a function of input fidelity $f$ and number $n$ of purification 
rounds. (c) Off-diagonal contributions $G_0(f)$ (dashed curve) and $G_3(f, G_2)$ 
(solid curve) given by Eqs.~(\ref{off-diag0}) and (\ref{off-diag}), respectively.}
\label{fig4}
\end{center}
\end{figure}

Being transported from the entanglement distribution block into the cavity $C_1$
($C_3$), the atoms $1_A, 2_A$ ($1_B, 2_B$) couple simultaneously to the same
cavity mode and both laser beams $L_2$ ($L_5$) and $L_1$ ($L_4$) as displayed in
Fig.~\ref{fig3}(b). Assuming the non-zero cavity relaxation rate $\kappa$ associated
with $C_1$ ($C_3$), the evolution of the coupled atom-cavity-laser system is
governed by the master equation \cite{me}
\begin{eqnarray}
&& \dot{\rho} = - \frac{\im}{\hbar} \left[ H_S, \rho \right]
                + \frac{\kappa}{2} \left( 2 \, a \, \rho \, a^\dag
                - a\, a^\dag \rho - \rho \, a^\dag a \right) \equiv \mathcal{L} \, \rho \, ;
                \label{me} \qquad \\
&& \hspace{2cm} H_S = \frac{\hbar \, J_2}{2} \left( a + a^\dagger \right) 
                      \left( \sigma_1^X + \sigma_2^X \right), \label{ham-acl}
\end{eqnarray}
where $\rho$ is the density operator describing the state of the two
atoms together with the cavity mode, $\mathcal{L}$ is the Lindbladian 
superoperator that acts on the density operator,
$\sigma^X_i$ is the respective Pauli operator in the basis $\{
\ket{0}, \ket{1} \}$, and $J_2$ is the coupling between the atoms
inserted into the same cavity mode and subjected to the two 
laser beams. We show in Appendix~\ref{app1} that the above Hamiltonian 
is produced deterministically in our setup assuming both (i) the strong
driving regime of atoms and (ii) the large detuning limit for laser and
cavity fields.

The evolution of atoms $1_A, 2_A$ ($1_B, 2_B$) coupled to the field
of cavity $C_1$ ($C_3$) due to Eq.~(\ref{me}) is completely
determined by the exponent $e^{\mathcal{L} t}$ and the initial
state of the cavity and the atoms. Since the pairs of atoms $1_{AB}$ and
$2_{AB}$ are initially entangled, we have to consider the composite 
density operator
\begin{equation}\label{density0}
\tilde{\rho}_{AB} = e^{\left( \mathcal{L}_A + \mathcal{L}_B \right) t} 
                       \left( \rho_f^1 \otimes \rho_f^2 \otimes
                              \ket{\bar{0}_A, \bar{0}_B} \bra{\bar{0}_A, \bar{0}_B} \right) \, ,
\end{equation}
describing the state of two atomic pairs and two initially empty
cavities at a given time $t$. In this expression, $\rho_{f}^1$ and
$\rho_{f}^2$ are the density operators of the entangled pairs $1_{AB}$
and $2_{AB}$, respectively, while $\ket{\bar{0}_A}$ and
$\ket{\bar{0}_B}$ denote the vacuum states of cavities $C_1$ and
$C_3$, respectively. In Appendix~\ref{app2} we show, moreover, that
conditioned upon the no-photon measurement of the leaked cavity
field in both repeater nodes, the state (\ref{density0}) reduces to
the state (\ref{density3}) in the steady-state regime ($\kappa \, t \gg 1$), 
that is
\begin{equation}\label{proj1}
\rho_f^{1,2} = \frac{\bra{\bar{0}_A, \bar{0}_B} \tilde{\rho}_{AB}^{ss}
               \ket{\bar{0}_A, \bar{0}_B}}{\text{Tr} \left[
               \bra{\bar{0}_A, \bar{0}_B} \tilde{\rho}_{AB}^{ss}
               \ket{\bar{0}_A, \bar{0}_B} \right]} \, ,
\end{equation}
where $\tilde{\rho}_{AB}^{ss}$ is the operator (\ref{density0}) in the 
steady-state regime. The measurement of the leaked cavity field is performed 
using the photon detector $D_1$ ($D_2$) that is connected to the neighboring 
repeater node through a classical communication channel. We stress that since 
the detection of the leaked cavity field in our scheme discriminates between 
a vacuum state (no clicks) and a strong coherent state (many clicks), the 
efficiency of the detectors $D_1$ and $D_2$ can take rather moderate values.

Assuming that the four-qubit entangled state (\ref{density3}) has been successfully
generated, the permanent pair $P_{AB}$ is supplemented by two temporary pairs $1_{AB}$
and $2_{AB}$ as displayed in Fig.~\ref{fig3}(a). Similar to the original scheme, each
repeater node contains now one triplet of qubits and each of these triplets evolves
due to the isotropic Heisenberg XY Hamiltonian
\begin{equation}\label{ham-xy1}
H_{XY} = \frac{\hbar \, J_3}{2} \sum_{i = 1}^3
         \left( \sigma_i^X \sigma_{i+1}^X + \sigma_i^Y \sigma_{i+1}^Y \right),
\end{equation}
over the time period ($n = 0,1,2, \ldots$)
\begin{equation}\label{time1}
T = \frac{\pi}{3} \left( n + \frac{1}{2} \right) J_3^{-1} \, ,
\end{equation}
such that $\sigma_4^X = \sigma_1^X$ and $\sigma_4^Y = \sigma_1^Y$, and where $J_3$
is the coupling between the qubits. In Appendix~\ref{app3}, we show that the
Hamiltonian (\ref{ham-xy1}) is produced deterministically in our scheme by
coupling simultaneously three atoms to the same cavity mode $C_2$ ($C_4$)
and the laser beam $L_1$ ($L_4$) in the large detuning limit [see Fig.~\ref{fig3}(c)].

Similar to the original scheme, this evolution is followed by the projective
measurement of qubit pairs $1_{AB}$ and $2_{AB}$ in the basis $\{ \ket{0}, \ket{1} \}$
and the exchange of the projection outcomes between the two repeater nodes by means of 
classical communication. The entanglement purification is successful if the outcome of 
projections agrees with (\ref{outcome}). In this case, the (unprojected) permanent 
qubit pair is described again by the density operator (\ref{density2}), where
\begin{equation}\label{final4}
F(f, f^\prime) = \frac{(25 - 50 f + 194 f^2) f^\prime}
                 {169 + 194 f^2 - 144 f^\prime + (288 f^\prime - 338) f} \, ,
\end{equation}
such that $\mathbf{F}(\rho_F) = F(f, f^\prime) > \mathbf{F}(\rho_{f^\prime})$.
In Fig.~\ref{fig4}(a), we compare the fidelity
\begin{equation}\label{final5}
F(f, f) = \frac{(25 - 50 f + 194 f^2) f} {169 - 482 f + 482 f^2}
\end{equation}
(solid curve) with the respective fidelity given by
Eq.~(\ref{final1}) (dotted curve). We see that the growth of
fidelity in the modified scheme is almost twice as large as in the
original purification scheme. This nice result, however, relies
merely on the input state (\ref{density3}) that is entangled strongly
if compared to the separable state $\rho_f^1 \otimes \rho_f^2$ used in
the original scheme. We recall that our scheme relies on the effect
of entanglement transfer between the networks of evolving spin
chains introduced in Refs.~\cite{qip, pra78a} and realized in our scheme
using the cavity QED framework. The stronger the entangled state 
that we provide as the input for the purification block in our scheme,
the more entanglement is transferred to the permanent qubit pair.

Assuming that each purification round is successful, the sequence
from Fig.~\ref{fig3}(a) leads to a gradual growth of entanglement
fidelity (of stationary atoms) with regard to the respective
fidelity obtained in the previous round. Similar to the original
scheme, we analyze quantitatively the sequence (\ref{seq4}) in order
to understand how much the output fidelity increases with each
purification round. In Fig.~\ref{fig4}(b), we show the plot of
$\widehat{F}(f, n)$ that describes the difference between the final
fidelity $F_{n}(f, F_{n-1})$ obtained after $n$ (successful) rounds
and the initial fidelity $f$ ($n = 0$). For $f > 0.75$, this
function exhibits a dramatic growth during the first three rounds
that saturates and, with increasing $n$, yields a negligible
growth with regard to the following fixed point fidelity
\begin{widetext}
\begin{equation}\label{final6}
F_{3}(f, F_{2}) = \frac{f \left( 25 - 50 f + 194 f^2 \right)^3} {4826809 - 33772038 f
                  + 103411314 f^2 - 179097440 f^3 + 189095940 f^4 - 119456664 f^5 + 39818888 f^6} \, ,
\end{equation}
\end{widetext}
displayed in Fig.~\ref{fig4}(a) by a dashed curve. We readily see that we obtain an almost unit
output fidelity for $f > 0.75$ after the same numner of purification rounds as in the original
scheme. For $f < 0.75$, however, the function $\widehat{F}(f, n)$ continues to grow with each
purification round. In this case, the optimal number of purification rounds and the respective
fixed point fidelity has to be determined for each particular value of $f$ separately.

\subsection{Evolution governed by Hamiltonian (\ref{ham-xy1}) and the purification gate}

In this section, we already explained that the density operator (\ref{density2}) 
with the function (\ref{final4}) characterize completely the permanent qubit pair obtained in
our (modified) scheme after a single purification round. In this subsection, we analyze briefly
the evolution governed by the Hamiltonian (\ref{ham-xy1}) and connect it with the main results
utilized in this paper.

The atomic evolution governed by Hamiltonian (\ref{ham-xy1})
\begin{equation}\label{evol1}
e^{- \frac{\im}{\hbar} H_{XY} \, t} = \sum_{k=1}^8 e^{- \frac{\im}{\hbar} E_k \, t} \, \ket{k} \bra{k} \, ,
\end{equation}
is completely determined by the energies $E_k$ and vectors $\ket{k}$, which satisfy the
eigenvalue equality $H_{XY} \, \ket{k} = E_k \, \ket{k}$
with orthogonality and completeness relations $\braket{k}{k^\prime} = \delta_{k k^\prime}$
and $\sum \ket{k} \bra{k} = I$, respectively. With the help of Jordan-Wigner transformation
\cite{zp47}, this eigenvalue problem can be solved exactly (see, for instance, Ref.~\cite{pra64}).
Since the evolution operator (\ref{evol1}) acts on the states of one atomic triplet that is
entangled with another atomic triplet in the neighboring node, we have to consider the
composite evolution operator
\begin{equation}\label{evol2}
U(t) = \sum_{k,k^\prime =1}^8 e^{- \frac{\im}{\hbar} \left( E_k + E_{k^\prime} \right) \, t}
          \, \ket{k_A \otimes k_B^\prime} \bra{k_A \otimes k_B^\prime} \, .
\end{equation}

Earlier we explained that right before each atomic pair from node A enters the cavity, it
becomes entangled with another atomic pair from node B, such that the four-qubit state
(\ref{density3}) is generated. We denote the density operator of the stationary atoms by
$\rho_{f^\prime}^3$. According to the evolution operator (\ref{evol2}) and this notation,
the state of both atomic triplets in nodes A and B is described by the six-qubit density
operator
\begin{equation}\label{density4}
\rho^{1,2,3}(t, f, f^\prime) = U(t) \left( \rho_{f}^{1,2} \otimes \rho_{f^\prime}^3 \right) U^\dag(t) \,
\end{equation}
that evolves over the time period $T$ given by Eq.~(\ref{time1}). After this evolution,
the state of both atomic triplets in nodes A and B is described by the density operator
\begin{equation}\label{density5}
\rho^{1,2,3}(T, f, f^\prime) =
     \sum_{i,j=1}^{64} \, \rho^{1,2,3}_{ij}(T, f, f^\prime) \, \ketv{i} \brav{j},
\end{equation}
where $2^6$ composite vectors $\ketv{i}$, satisfying the orthogonality and completeness
relations $\braketv{i}{j} = \delta_{ij}$ and $\sum \ketv{i} \brav{i} = I$, respectively,
have been introduced.

In order to finalize one purification round, the conveyed atomic pairs are projectively
measured, such that the projected density operator
\begin{equation}\label{density6}
\rho (T, f, f^\prime) = \sum_{\alpha, \beta = 1}^4
      \frac{ \rho^{1,2,3}_{\alpha \beta}(T, f, f^\prime)}
      {P_{\text{succ}}(T, f, f^\prime)} \, \ketvt{\alpha} \bravt{\beta},
\end{equation}
with      
\begin{equation}
P_{\text{succ}}(T, f, f^\prime) =
      \text{Tr} \left[ \sum_{\alpha, \beta = 1}^4
      \rho^{1,2,3}_{\alpha \beta}(T, f, f^\prime)  \,
      \ketvt{\alpha} \bravt{\beta} \right], \notag
\end{equation}
describes the state of the stationary atoms. In the above expressions, the Greek
indices run over the four different values given by
\begin{subequations}
\begin{eqnarray}
\ketvt{\alpha} \equiv \langle 0_{1A}, 1_{2A}, 0_{1B}, 1_{2B} | \mathbf{v}_\alpha \rangle \neq 0 \, , && \\
\text{or} \quad
\ketvt{\alpha} \equiv \langle 1_{1A}, 0_{2A}, 1_{1B}, 0_{2B} | \mathbf{v}_\alpha \rangle \neq 0 \, , &&
\end{eqnarray}
\end{subequations}
which correspond to the outcomes of the projections (\ref{outcome}).

Using the six-qubit density operator (\ref{density4}), we have routinely computed
the matrix elements $\rho^{1,2,3}_{ij}(T, f, f^\prime)$ which, however, are rather
bulky to be displayed here. With the help of these matrix elements, we confirmed
that the density operator (\ref{density6}) coincides with the rank-two mixed state
(\ref{density2}), where the function $F(f, f^\prime)$ is given by Eq.~(\ref{final4}).

\subsection{Remarks on the entanglement distribution between stationary atomic qubits}

Throughout the paper, we assumed that the stationary atoms are initially entangled, 
such that the fidelity is above the threshold value of $1/2$. In our previous paper, 
we suggested that there is no need to introduce an additional entanglement distribution 
protocol in our setup in order to entangle the stationary atoms prior to the purification. 
Instead, it was suggested to prepare initially both permanent atoms in the ground 
state and run our purification scheme. We showed that one successful purification round 
\textit{induces} the entanglement of the stationary atoms and ensures that the fidelity of 
resulting density operator (almost) coincides with the fidelity $f$ of the temporary pairs 
$\rho_f^1$ and $\rho_f^2$. In other words, one purification round entangles two (initially 
separable) stationary atoms, such that the fidelity of the temporary pairs is mapped to 
the fidelity of the stationary density operator.

In our modified scheme, we utilize the same procedure as in the original scheme. We 
generate two (separate) entangled pairs and let them be conveyed through the 
cavity $C_1$ ($C_3$), such that the (probabilistic) entanglement protocol that produces 
the four-qubit state (\ref{density3}) is switched off. It can be shown that a successful 
purification round with these two entangled pairs transforms the state of the permanent 
atoms (prepared initially in the ground state) into an entangled state described by
\begin{eqnarray}\label{state}
\rho_f &=& F_0(f) \, \Phi^+_{A,B} + \left( 1 - F_0(f) \right) \Phi^-_{A,B} \notag \\
       &+& G_0(f) \left( \ket{\phi^+_{A,B}} \bra{\phi^-_{A,B}} + 
                             \ket{\phi^-_{A,B}} \bra{\phi^+_{A,B}} \right) \, ,
\end{eqnarray}
with
\begin{equation}\label{off-diag0}
F_0(f) = \frac{1 + 48 f + 32 f^2}{\sum_i (-1)^i \, c_i \, f^i} \, ; \quad
G_0(f) = \frac{9 - 32 f + 32 f^2}{\sum_i (-1)^i \, c_i \, f^i} \, ,
\end{equation}
where $c_0 = 82$, $c_1 = c_2 = 64$ are the only non-zero coefficients.

Obviously, the above state is no longer a rank-two mixed state like in 
Eq.~(\ref{density1}) because of the off-diagonal contribution $G_0(f)$ displayed 
in Fig.~\ref{fig4}(c) by a dashed curve. The fidelity (\ref{f-def}) associated with 
(\ref{state}), however, is slightly larger than the fidelity $f$ associated with the 
temporary pairs $\rho_f^1$ and $\rho_f^2$. The role of this \textit{initialization} 
round is solely to entangle the stationary atoms and, therefore, it has to be followed 
by a number of purification rounds leading to the gradual growth of fidelity,
\begin{equation}\label{seq5}
F_{0}(f) < F_{1}(f, F_{0}) < \ldots < F_{n}(f, F_{n-1}) \, ,
\end{equation}
and the gradual reduction of (off-diagonal) contributions, 
\begin{equation}\label{seq6}
G_{0}(f) > G_{1}(f, G_{0}) > \ldots > G_{n}(f, G_{n-1}) \, .
\end{equation}

Using the above sequences, we calculated the fidelity 
\begin{equation}\label{final7}
F_{3}(f, F_{2}) = \frac{(1 + 48 f + 32 f^2)(25 - 50 f + 194 f^2)^3}{2 \, \sum_i (-1)^i \, d_i \, f^i} \, ,
\end{equation}
and the respective off-diagonal contribution
\begin{equation}\label{off-diag}
G_3(f, G_2) = \frac{274625 \, (9 - 32 f + 32 f^2) (1 - 2 f + 2 f^2)^3}{2 \, \sum_i (-1)^i \, d_i \, f^i} \, ,
\end{equation}
which are motivated by the optimal number of purification rounds obtained previously
[see (\ref{final3}) and (\ref{final6})], and where 
\begin{eqnarray}
&& d_0 = 195493577; \ d_1 = 1442887766; \ d_2 = 4716352898; \notag \\
&& d_3 = 8883640864; \ d_4 = 10517241220; \ d_5 = 7944708952; \notag \\
&& d_6 = 3738576328; \ d_7 = 934577152; \ d_8 = 233644288 \notag
\end{eqnarray}
are the only non-zero coefficients.

In contrast to the dashed curve describing $G_{0}(f)$ in Fig.~\ref{fig4}(c), the 
solid curve describing (\ref{off-diag}) deviates slightly around 
the constant value of $0.004$. To a good approximation, therefore, the off-diagonal 
contribution $G_3(f, G_2)$ can be neglected and the resulting density operator takes 
the form of the rank-two mixed state (\ref{density1}). We have verified, moreover, that 
the output fidelity (\ref{final7}) (almost) coincides with the output fidelity 
($\ref{final6}$) displayed in Fig.~\ref{fig4}(a) by a dashed curve. The price we 
pay for one extra (successful) purification round prior to the main sequence of rounds, 
therefore, is clearly compensated by the more moderate demand of physical resources in 
our purification scheme.

\section{Summary and discussion}

In this paper, an efficient, high-fidelity scheme was proposed to purify 
the low-fidelity entangled atoms trapped in two remote optical 
cavities. This scheme is a modification of the purification scheme proposed in 
our previous paper \cite{pra84} that exploits the natural evolution of spin 
chains instead of CNOT gates. Similar to the original scheme, the modified 
scheme uses a cavity-QED framework, namely (i) short chains of atoms, 
(ii) high-finesse optical cavities, and (iii) detectors for the projective
measurement of atomic states. In contrast to the original scheme, however,
one additional entanglement protocol was introduced in each repeater 
node, and the laser beams which are used to control the entire scheme were optimized.
With the help of these modifications, an almost unit output fidelity was achieved
after the same number of successful purification rounds as in the original scheme.
Similar to the original paper, furthermore, the modified scheme was supplied with a 
detailed experimental setup, and a complete description of all necessary steps and 
manipulations has been given. A comprehensive analysis of fidelities obtained after 
multiple purification rounds was performed and the optimal number of rounds was 
determined. We also discussed in detail the initial distribution of entanglement 
between the stationary qubits trapped in two remote cavities.

Throughout the paper, we assumed that each purification round is finalized successfully 
leading to a gradual growth (\ref{seq3}) or (\ref{seq5}) of entanglement fidelity.
In the case of an unsuccessful purification event, i.e., when the outcomes of the projective 
measurement disagree with (\ref{outcome}), the stationary atoms should be reinitialized 
and the entire scheme restarted. Since the probability to get the two (out of $2^4$) 
combinations of projective measurements for a successful purification is rather small, 
the occurrence of multiple unsuccessful events can require a large amount of atomic pairs 
in the chain and unreasonable operational 
times. We stress, therefore, that although the proposed purification scheme is experimentally 
feasible, a practical mechanism that reduces unsuccessful purification events has to
be considered. This problem and possible solutions shall be addressed in our future work.

The high-fidelity purification scheme proposed in this paper enables one to perform 
multiple entanglement swapping operations and thus opens a route towards an efficient 
and experimentally feasible quantum repeater for long-distance quantum communication. 
More specifically, in our experimental setup,
each atom in node A has to be entangled with another atom from node B right 
before they enter the cavities $C_1$ and $C_3$ for further processing. The (low-fidelity) 
entanglement between these atoms is distributed non-locally using the entanglement 
distribution block indicated in Fig.~\ref{fig3}(d) by a rectangle. In order to
entangle two (three-level) atoms located at distant repeater nodes A and B, 
we find the entanglement distribution scheme proposed in Ref.~\cite{prl96a} the most
appropriate. This scheme is also realizable in the framework of cavity-QED and, 
therefore, it utilizes the same physical resources as our purification scheme.

By this scheme, a coherent-state light pulse interacts with the coupled atom-cavity 
system in node A, such that the optical field accumulates a phase conditioned upon the 
atomic state in this node. Afterwards, the light pulse propagates to node B, 
where it interacts with the second coupled atom-cavity system and accumulates another 
phase conditioned upon the atomic state in this node. The resulting density operator \cite{pra78b}
\begin{equation}\label{mess}
f \, \ket{\widetilde{\phi}^+_{A,B}} \bra{\widetilde{\phi}^+_{A,B}} 
       + (1 - f) \ket{\widetilde{\phi}^-_{A,B}} \bra{\widetilde{\phi}^-_{A,B}} \, ,
\end{equation}        
with
\begin{eqnarray}
\ket{\widetilde{\phi}^\pm_{A,B}} &\equiv& \frac{1}{\sqrt{2}} \ket{C_0} \ket{\phi^\pm_{A,B}} \notag \\
		&\pm& \frac{1}{2} \, e^{- \im \eta \epsilon} \ket{C_1} \ket{1_A, 0_B} 
		  + \frac{1}{2} \, e^{\im \eta \epsilon} \ket{C_2} \ket{0_A, 1_B} \, , \notag
\end{eqnarray}
describes the state of both atoms and the coherent light pulse, where $C_0 \equiv \sqrt{\eta} \, \alpha$, 
$C_1 \equiv \sqrt{\eta} \, \alpha \, e^{\im \theta}$, $C_2 \equiv \sqrt{\eta} \, \alpha \, e^{- \im \theta}$
denote the phase-rotated and channel-damped coherent state $\alpha$, $\epsilon \equiv \alpha^2 \sin \theta$, 
while $f \equiv ( 1 + e^{-(1 - \eta) \alpha^2 (1 - \cos \theta)} )/2$ plays the role of the entanglement
fidelity.

The resulting (phase-rotated) coherent pulse becomes disentangled from the atoms with the help of 
homodyne detection followed by post-selection \cite{prl96a} or, alternatively, using unambiguous state 
discrimination \cite{pra78b}. This projects the state (\ref{mess}) 
onto an entangled state of two atoms that coincides with the rank-two mixed state (\ref{density1}). 
We remark that the conditioned phase rotation exploited in the entanglement distribution scheme 
is naturally realized in a cavity-QED framework using the single atom-cavity evolution in the 
dispersive interaction regime.

\begin{acknowledgments}

We thank the DFG for support through the Emmy Noether program.
In addition, we thank the BMBF for support through the QuOReP program.

\end{acknowledgments}

\appendix

\section{Derivation of the Hamiltonian (\ref{ham-acl})}\label{app1}

In this appendix, we show that the Hamiltonian (\ref{ham-acl}) is produced 
deterministically in our setup. Specifically, two (three-level) atoms are 
subjected to the field of the (initially empty) cavity $C_1$ ($C_3$) and the 
fields of laser beams $L_1$ ($L_4$) and $L_2$ ($L_5$) simultaneously as 
displayed in Fig.~\ref{fig3}(b). The evolution of this coupled atom-cavity-laser 
system is governed by the Hamiltonian ($k=1,2$)
\begin{eqnarray}\label{ham1}
H_1 &=& \hbar \, \omega_C \, a^\dag \, a - \im \hbar \sum_{k} 
       \left[ \frac{g}{2} \, a \, \ket{e}_k \bra{0} \right. \\
       &+& \left. \frac{\Omega}{2} \left(
       e^{-i \omega_L \, t} \ket{e}_k \bra{1} +
       e^{-i \omega_P \, t} \ket{e}_k \bra{0}\right) - H.c. \right] \notag \\
       &+& \hbar \sum_{k} \left[
       \omega_1 \ket{1}_k \bra{1} +
       \omega_E \ket{e}_k \bra{e} +
       \omega_0 \ket{0}_k \bra{0} \right] \, , \notag
\end{eqnarray}
where $g$ denotes the coupling strength of an atom to the cavity mode, 
while $\Omega$ denotes the coupling strengths of an atom to both laser 
fields.

We assume that $\omega_C = \omega_P$ and switch to the interaction picture using 
the unitary transformation
\begin{equation}\label{picture1}
U_1 =    e^{- \im t \left[\sum \left(
         \omega_1 \ket{1}_k \bra{1} + \omega_E \ket{e}_k \bra{e}
         + (\omega_L + \omega_1 - \omega_P ) \ket{0}_k \bra{0} \right) 
         + \omega_P \, a^\dag a \right]}. \notag
\end{equation}
We assume, moreover, that $\Delta_L = \Delta_C \equiv - \Delta$, where the notation 
$\Delta_L \equiv (\omega_E - \omega_1) - \omega_L$ and 
$\Delta_C \equiv (\omega_E - \omega_0) - \omega_C$ has been introduced. In the above
interaction picture, therefore, the Hamiltonian (\ref{ham1}) takes the following form
\begin{eqnarray}\label{ham2}
H_2 &=& - \im \hbar \sum_{k} 
       \left[ \frac{g}{2} \, e^{-i \Delta \, t} a \, \ket{e}_k \bra{0} \right. \\
    &+& \left. \frac{\Omega}{2} \, e^{-i \Delta \, t} \left(
       \ket{e}_k \bra{1} + \ket{e}_k \bra{0}\right) - H.c. \right] . \notag 
\end{eqnarray}

We require that $\Delta$ is sufficiently far detuned, such that no atomic 
$\ket{e} \leftrightarrow \ket{0}$ and $\ket{e} \leftrightarrow \ket{1}$ 
transitions can occur. We expand the evolution governed by the 
Hamiltonian (\ref{ham2}) in series and keep the terms up to the second order,
\begin{equation}
U_2 \cong \vecI - \frac{\im}{\hbar} \int_{0}^{t} H_2 \, dt^\prime - 
					\frac{1}{\hbar^2} \int_{0}^{t} \left( H_2 \, \int_{0}^{t^\prime} 
					H_2 \, dt^{\prime \prime} \right) dt^\prime \, . \notag
\end{equation}
By performing integration and retaining only linear-in-time contributions, 
we express this evolution in the form
\begin{equation}\label{operator2}
U_2 \cong \vecI - \frac{\im}{\hbar} \, H_3 \, t 
    \cong \exp \left[ - \frac{\im}{\hbar} \, H_3 \, t \right],
\end{equation}
where the effective Hamiltonian is given by
\begin{equation}\label{ham3}
H_3 = \frac{\hbar \, \Omega}{4 \Delta} \sum_{k} 
      \left[ \Omega \, \ket{1}_k \bra{0} + g \, \ket{1}_k \bra{0} \, a + H.c. \right] \, ,
\end{equation}
after removing the constant contributions.

At this stage, we switch from the atomic basis $\{ \ket{0}, \ket{1} \}$ 
to the basis $\{ \ket{+}, \ket{-} \}$, where
\begin{equation}\label{basis}
\ket{+} = \frac{1}{\sqrt{2}} \left( \ket{0} + \ket{1} \right);  \quad 
\ket{-} = \frac{1}{\sqrt{2}} \left( \ket{0} - \ket{1} \right).
\end{equation}
In this basis, the Hamiltonian (\ref{ham3}) takes the form
\begin{eqnarray}\label{ham4}
H_4 &=& \frac{\hbar \, \Omega}{8 \Delta} \sum_{k} \left[ 2 \, \Omega S^Z_k \right. \\
    && \left. \qquad + g \left( S^Z_k ( a + a^\dag ) + (S^\dag_k - S_k)(a - a^\dag)\right) 
     \right] , \notag
\end{eqnarray}
where $S_k \equiv \ket{-}_k \bra{+}$ and $S^Z_k \equiv \ket{+}_k \bra{+} - \ket{-}_k \bra{-}$,
and where we removed any constant contributions. We switch once more to the 
interaction picture with respect to the first term of (\ref{ham4}). In this 
interaction picture, we obtain
\begin{eqnarray}\label{ham5}
H_5 &=& \hbar \, \frac{g \, \Omega}{8 \Delta} \sum_{k} \left[ S^Z_k ( a + a^\dag ) \right. \\
    && \left. \qquad + (S^\dag_k \, e^{\im \frac{\Omega^2}{2 \Delta} t} 
                      - S_k \, e^{-\im \frac{\Omega^2}{2 \Delta} t})(a - a^\dag) \right] . \notag
\end{eqnarray}

In the strong driving regime, i.e., for $\Omega \gg \{ g, \Delta \}$, we eliminate 
the last (fast oscillating) term using the same arguments as for the rotating wave
approximation. The Hamiltonian (\ref{ham5}), therefore, reduces to
\begin{equation}\label{ham6}
H_6 = \hbar \, \frac{g \, \Omega}{8 \Delta} \left( a + a^\dag \right) 
                                            \left( \sigma^X_1 + \sigma^X_2 \right) \, ,
\end{equation}
where we used the identity $S^Z_k = \sigma^X_k$. The resulting Hamiltonian (\ref{ham6})
coincides with the Hamiltonian (\ref{ham-acl}) under the notation 
$J_2 \equiv g \, \Omega / (4 \Delta)$.

\section{Steady-state solution of Eq.~(\ref{me})}\label{app2}

In this appendix, we show that the mixed state (\ref{density3}) is conditionally
generated by means of evolution (\ref{me}) that takes place simultaneously in  
both repeater nodes (A and B) with the initial state 
\begin{equation}\label{density-i}
\tilde{\rho}^0_{AB} = \rho_f^1 \otimes \rho_f^2 \otimes
                           \ket{\bar{0}_A, \bar{0}_B} \bra{\bar{0}_A, \bar{0}_B}
\end{equation}
in the steady-state regime. In Sec.~III, we considered the expression (\ref{density0}) 
based on the exponent $e^{\mathcal{L} t}$ that is difficult to evaluate. Here we apply 
sequentially the steady-state solution of Eq.~(\ref{me}) to the coupled (atom-atom-cavity) 
system in node A and, afterwards, to the coupled system in node B.

In order to proceed, we consider first the solution of (\ref{me}) for an initially empty 
cavity $C_1$ and a general mixed state of two atoms in node A ($i,j = 1, \ldots ,4$)
\begin{equation}\label{density-}
\rho^0_A = \sum_{i,j} (\rho^0_A)_{ij} \, \ketu{i}{A} \brau{j}{A}
           \otimes \ket{\bar{0}_A} \bra{\bar{0}_A} \, ,
\end{equation}
where we switched to the atomic basis (\ref{basis}), such that 
$\ketu{1}{A} = \ket{+_{1 A}, +_{2 A}}$, $\ketu{2}{A} = \ket{-_{1 A}, -_{2 A}}$, 
$\ketu{3}{A} = \ket{+_{1 A}, -_{2 A}}$, and $\ketu{4}{A} = \ket{-_{1 A}, +_{2 A}}$ form 
together an orthogonal basis. To our best knowledge, 
the master equation (\ref{me}) was solved in Refs.~\cite{bina, loug} only for an initial 
pure state of atoms, which is not appropriate to be used in our case. We therefore (re)solved 
this master equation for an initial mixed state of two atoms (\ref{density-}). Assuming 
the strong atom-cavity coupling $J_2 \gg \kappa$, the solution we found in the steady-state 
regime $\kappa \, t \gg 1$ can be expressed as
\begin{equation}\label{density-0}
\rho^{ss}_A = \sum_{i,j} (\rho^0_A)_{ij} \, \lambda_{ij} \, \ketu{i}{A} \brau{j}{A} 
            \otimes \ket{- u_i \, \alpha^{ss}_A} \bra{- u_j \, \alpha^{ss}_A} \, ,
\end{equation}
where $\alpha^{ss}_A = 2 \, \im J_2 / \kappa$ is the amplitude of the coherent state, 
$\lambda_{ij} = \delta_{(u_i - u_j),0}$ with $u_1 = 1$, $u_2 = -1$, and $u_3 = u_4 = 0$.

Recall that the evolution (\ref{me}) takes place simultaneously in both repeater 
nodes A and B with the initial state (\ref{density-i}), which we cast into the form 
\begin{eqnarray}\label{density-i1}
\tilde{\rho}^0_{AB} &=& \sum_{i,j,k,l} (\tilde{\rho}^0_{AB})_{ijkl} \, 
 											  \ket{\mathbf{u}_i^A, \mathbf{u}_k^B} \bra{\mathbf{u}_j^A, \mathbf{u}_l^B}
 											  \otimes  \ket{\bar{0}_A, \bar{0}_B} \bra{\bar{0}_A, \bar{0}_B} \notag \\
 										&\equiv&	\sum_{i,j} (\rho^0_A)_{ij} \, \ketu{i}{A} \brau{j}{A}
           							\otimes \ket{\bar{0}_A} \bra{\bar{0}_A} \, , 
\end{eqnarray}
where we introduced the matrices
\begin{eqnarray}
(\tilde{\rho}^0_{AB})_{ijkl} &=& \bra{\mathbf{u}_i^A, \mathbf{u}_k^B} \, 
                                 \rho_f^1 \otimes \rho_f^2 \, \ket{\mathbf{u}_j^A, \mathbf{u}_l^B} \, , \\
(\rho^0_A)_{ij} &=& \sum_{k,l} (\tilde{\rho}^0_{AB})_{ijkl} \, \ketu{k}{B} \brau{l}{B} 
                               \otimes \ket{\bar{0}_B} \bra{\bar{0}_B} \, . \quad
\end{eqnarray}

By identifying the expressions (\ref{density-})  and (\ref{density-i1}), we conclude 
that the density operator (\ref{density-0}) gives the steady-state solution of (\ref{me}) 
obtained in node A for the initial state (\ref{density-i}). Conditioned upon the 
no-photon measurement of the leaked field from the cavity $C_1$, this steady-state 
solution reduces to
\begin{eqnarray}\label{density-1}
\bra{\bar{0}_A} \tilde{\rho}^{ss}_{A} \ket{\bar{0}_A} &=& \sum_{i,j} (\rho^0_A)_{ij} \, \lambda_{ij} \, 
                                                   e^{-|\alpha^{ss}_A|^2 u_i u_j} \ketu{i}{A} \brau{j}{A} 
                                                   \qquad \notag \\
                                               &\equiv& \sum_{k,l} (\rho^0_B)_{kl} \, \ketu{k}{B} \brau{l}{B}
           												      					 \otimes \ket{\bar{0}_B} \bra{\bar{0}_B} \, ,    
\end{eqnarray}
where
\begin{equation}\label{density-5}
(\rho^0_B)_{kl} \equiv \sum_{i,j} (\tilde{\rho}^0_{AB})_{ijkl} \, 
                  \lambda_{ij} \, e^{-|\alpha^{ss}_A|^2 u_i u_j} \ketu{i}{A} \brau{j}{A} \, .
\end{equation}

Similar to node A, the density operator 
\begin{equation}\label{density-6}
\rho^{ss}_B = \sum_{i,j} (\rho^0_B)_{ij} \, \lambda_{ij} \, \ketu{i}{B} \brau{j}{B} 
            \otimes \ket{- u_i \, \alpha^{ss}_B} \bra{- u_j \, \alpha^{ss}_B} \, ,
\end{equation}
gives the steady-state solution of (\ref{me}) for node B with the initial state (\ref{density-5}). 
Conditioned upon the no-photon measurement of the leaked field from the cavity $C_3$, this solution 
reduces to
\begin{eqnarray}\label{density-2}
\bra{\bar{0}_A, && \bar{0}_B} \tilde{\rho}^{ss}_{AB} \ket{\bar{0}_A, \bar{0}_B} \notag \\
                  && = \sum_{k,l} \lambda_{kl} (\rho^0_B)_{kl} \, e^{-|\alpha^{ss}_B|^2 u_k u_l} \, 
                    \ketu{k}{B} \brau{l}{B} \notag \\
                  && = \sum_{i,j,k,l} (\tilde{\rho}^0 \lambda)_{ijkl} \, e^{-|\alpha^{ss}|^2 \theta_{ijkl}} 
                   \ket{\mathbf{u}_i^A, \mathbf{u}_k^B} \bra{\mathbf{u}_j^A, \mathbf{u}_l^B} \, , \qquad
\end{eqnarray}
where the notation $\theta_{ijkl} = u_i u_j + u_k u_l$, $(\tilde{\rho}^0 \lambda)_{ijkl} = 
(\tilde{\rho}^0_{AB})_{ijkl} \, \lambda_{ij} \lambda_{kl}$, and 
$\alpha^{ss}_A = \alpha^{ss}_B \equiv \alpha^{ss}$ was introduced. The atom-cavity strong 
coupling ensures that $|\alpha^{ss}|^2 \gg 1$. To a good approximation, this observation  
implies that the exponent in (\ref{density-2}) vanishes for $\theta_{ijkl} \neq 0$
\begin{equation}\label{density-3}
\bra{\bar{0}_A, \bar{0}_B} \tilde{\rho}^{ss}_{AB} \ket{\bar{0}_A, \bar{0}_B} 
             = \sum_{i,j,k,l}^{\theta_{ijkl} = 0} (\tilde{\rho}^0 \lambda)_{ijkl} 
               \ket{\mathbf{u}_i^A, \mathbf{u}_k^B} \bra{\mathbf{u}_j^A, \mathbf{u}_l^B} \, .
\end{equation}

Owing to the explicit form of (\ref{density-i}), we have routinely computed the matrix 
elements $(\tilde{\rho}^0_{AB})_{ijkl}$ which, however, are rather bulky to be displayed 
here. With the help of these matrix elements, we obtained the normalized density operator 
(\ref{density-3}) in the form
\begin{widetext}
\begin{eqnarray}\label{density-4}
\rho_f^{1,2} = \frac{f^2}{2 - 4 f + 4 f^2} && \left( \ket{\phi^+_{1A,1B}, \phi^+_{2A,2B}}
               \bra{\phi^+_{1A,1B}, \phi^+_{2A,2B}} + \ket{\psi^+_{1A,1B}, \psi^+_{2A,2B}}
               \bra{\psi^+_{1A,1B}, \psi^+_{2A,2B}} \right. \notag \\
               && \left. - \ket{\psi^+_{1A,1B}, \psi^+_{2A,2B}}
               \bra{\phi^+_{1A,1B}, \phi^+_{2A,2B}} - \ket{\phi^+_{1A,1B}, \phi^+_{2A,2B}}
               \bra{\psi^+_{1A,1B}, \psi^+_{2A,2B}} \right) \notag \\
             + \frac{(f-1)^2}{2 - 4 f + 4 f^2} && \left( \ket{\phi^-_{1A,1B}, \phi^-_{2A,2B}}
               \bra{\phi^-_{1A,1B}, \phi^-_{2A,2B}} + \ket{\psi^-_{1A,1B}, \psi^-_{2A,2B}}
               \bra{\psi^-_{1A,1B}, \psi^-_{2A,2B}} \right. \notag \\
               && \left. - \ket{\psi^-_{1A,1B}, \psi^-_{2A,2B}}
               \bra{\phi^-_{1A,1B}, \phi^-_{2A,2B}} - \ket{\phi^-_{1A,1B}, \phi^-_{2A,2B}}
               \bra{\psi^-_{1A,1B}, \psi^-_{2A,2B}} \right)
\end{eqnarray}
\end{widetext}
that describes a four-qubit entangled state and coincides with the state (\ref{density3}).

\section{Derivation of the Hamiltonian (\ref{ham-xy1})}\label{app3}

In this appendix, we show that the Hamiltonian (\ref{ham-xy1}) is produced 
deterministically in our setup. Specifically, three (three-level) atoms are 
subject to the field of the (initially empty) cavity $C_2$ ($C_4$) and the 
field of laser beam $L_1$ ($L_4$) simultaneously as displayed in Fig.~\ref{fig3}(c). 
The evolution of this coupled atom-cavity-laser system is governed by the 
Hamiltonian ($k=1,2,3$)
\begin{eqnarray}\label{ham-1}
H_1 &=& \hbar \, \omega_C \, a^\dag \, a \\
       &-& \im \hbar \sum_{k} 
       \left[ \frac{g}{2} \, a \, \ket{e}_k \bra{0} + \frac{\Omega}{2} \,
       e^{-i \omega_L \, t} \ket{e}_k \bra{1} - H.c. \right] \notag \\
       &+& \hbar \sum_{k} \left[
       \omega_1 \ket{1}_k \bra{1} +
       \omega_E \ket{e}_k \bra{e} +
       \omega_0 \ket{0}_k \bra{0} \right] \, , \notag
\end{eqnarray}
where $g$ denotes the coupling strength of an atom to the cavity mode, 
while $\Omega$ denotes the coupling strength of an atom to the laser field.

We switch to the interaction picture using the unitary transformation
\begin{equation}\label{picture-1}
U_1 =    e^{- \im t \left[\sum \left(
         \omega_1 \ket{1}_k \bra{1} + \omega_E \ket{e}_k \bra{e}
         + \omega_0 \ket{0}_k \bra{0} \right) 
         + \left( \omega_1 + \omega_L - \omega_0 \right) a^\dag a \right]}. \notag
\end{equation}
In this picture, the Hamiltonian (\ref{ham-1}) takes the form
\begin{eqnarray}\label{ham-2}
H_2 &=& \hbar \, \left( \Delta_L - \Delta_C \right) \, a^\dag \, a \\
       &-& \im \hbar \sum_{k} 
       \left[ \frac{g}{2} \, a \, e^{i \Delta_L \, t} \ket{e}_k \bra{0} + \frac{\Omega}{2} \,
       e^{i \Delta_L \, t} \ket{e}_k \bra{1} - H.c. \right], \notag
\end{eqnarray}
where the notation $\Delta_L \equiv (\omega_E - \omega_1) - \omega_L$ and 
$\Delta_C \equiv (\omega_E - \omega_0) - \omega_C$ has been introduced.

We require that $\Delta_L$ and $\Delta_C$ are sufficiently far detuned, such that no atomic 
$\ket{e} \leftrightarrow \ket{0}$ and $\ket{e} \leftrightarrow \ket{1}$ transitions can occur. 
We expand the evolution governed by the Hamiltonian (\ref{ham-2}) in series up to the second 
order. By performing integration and retaining only linear-in-time contributions, we express 
this evolution in the form (\ref{operator2}), where the effective Hamiltonian is given by (we 
assume that the cavity field is initially in the vacuum state)
\begin{equation}\label{ham-3}
H_3 = \hbar \, \Delta \, a^\dag \, a 
      + \hbar \, \frac{g \, \Omega}{4 \, \Delta_L} \sum_{k} \left[ a \, \ket{1}_k \bra{0} + H.c. \right],
\end{equation}
where $\Delta \equiv \Delta_L - \Delta_C$. We switch one more time to the interaction picture 
with respect to the first term of (\ref{ham-3}). In this interaction picture, the Hamiltonian 
takes the form
\begin{equation}\label{ham-4}
H_4 = \hbar \, \frac{g \, \Omega}{4 \, \Delta_L} 
      \sum_{k} \left[ a \, e^{-i \Delta \, t} \ket{1}_k \bra{0} + H.c. \right].
\end{equation}

We require, finally, that $\Delta$ is sufficiently far detuned. As above, we expand again 
the evolution governed by the Hamiltonian (\ref{ham-4}) in series up to the second order and 
retain only linear-in-time contributions after the integration. This leads to the effective 
Hamiltonian ($i,j = 1,2,3$)
\begin{equation}\label{ham-5}
H_5 = \frac{\hbar \, g^2 \Omega^2}{16 \, \Delta_L^2 \, \Delta} 
      \left[\sum_{i,j}^{i \neq j} \ket{0_i, 1_j} \bra{1_i, 0_j} + \sum_k \ket{1}_k \bra{1} \right].
\end{equation}
Since the second term in this Hamiltonian commutes with the first term, we 
eliminate the second term by means of an appropriate interaction picture. The resulting 
Hamiltonian, i.e., the first term of (\ref{ham-5}), coincides with the Hamiltonian 
(\ref{ham-xy1}) under the notation $J_3 \equiv g^2 \Omega^2 / (16 \, \Delta_L^2 \, \Delta)$.

\end{document}